# A nanoionic diode: Equilibrium rectifying junction enabling large and stable resistance variations

Chuanlian Xiao*, Joachim Maier*

Max Planck Institute for Solid State Research, 70569, Stuttgart, Germany

*Corresponding authors. Email: c.xiao@fkf.mpg.de (C.X.); office-maier@fkf.mpg.de (J.M.)

ABSTRACT

We report on a new type of rectifier which is in full contact equilibrium and thus, if down-sized to the nanoscale, shows no drift even if exposed to elevated temperatures and/or extreme waiting times. This is in contrast to existing diodes which rely on frozen doping profiles and are hence non-equilibrium devices. Our rectifiers are related to Schottky diodes but employ "dopants" whose mobilities are high enough to follow the electrical field quickly but low enough to not compete with the electrons in terms of conductivities. In order to realize such a device based on mixed conductors, we use nanosized $TiO_2$ films on Ru as a substrate which can store Li at the interface according to a job-sharing mechanism (Li-ions on the $TiO_2$ side, electrons on the Ru side). The excellent functionality of this nanoionic device is demonstrated (e.g., current on-off ratio can exceed 6-7 orders of magnitude) and the additional advantages stressed (such as ease of preparation and tuning the characteristics electrochemically).

INTRODUCTION

Semiconductor diodes are indispensable elements of the present electronic technology, as they allow for efficient, fast and precise current rectification[1]. They are based on p-n junctions of two semiconductors or on Mott-Schottky layers at semiconductor-metal contacts. The latter are termed Schottky diodes. Both devices owe their function to appropriate doping. In the bulk, the seemingly immobile dopant is compensated by electrons or holes. If the latter are depleted in the space charge zones, one speaks of Mott-Schottky layers. The depletion effect can be substantially augmented on applying bias in the appropriate direction (blocking or reverse direction). Applying bias in the opposite direction decreases, virtually nullifies or even overcompensates the depletion effect (conducting or forward direction) leading to the rectifying effect of this type of diode. Schottky diodes are preferred over p-n junction rectifiers if fast switching or low voltage losses in forward direction are desired; they also exhibit typically lower threshold voltages, making them suitable for low-power operation. Of course, mobilities are finite for any ions, meaning that for a given bias, the dopant eventually tends towards an equilibrium distribution resulting in drift effects.

Such drift effects may become perceptible and critical if the devices are down-scaled to tiny dimensions or/and used at elevated temperatures. Note that not only the tendency to assume the equilibrium profile is the issue, also extremely minute variations directly at the interface can contribute to drift effects by modifying the barrier height. At higher temperatures, intrinsic carrier generation can become significant (owing to low band gap and low dopant solubilities) and disguise the doping effect[1-5]. Moreover, very high driving forces can increase the dopants' mobilities. Refs.[1,6-8] explicitly state that for Schottky diodes, the operation temperatures should not be above 150 °C[1,6-8] (see also Supplementary Note 1). For Ge-semiconductor systems with their typically greater dopant diffusivities, the safe operation temperatures should be even lower. In terms of high temperature application, SiC-based systems (instead of Si) have been recommended[4,8-10], for which, however, very high processing temperatures (> 1500 °C) are required.

For this reason, in the literature, efforts have been undertaken to identify *equilibrium* junctions which would function even under such critical conditions. Clearly, frozen doping profiles are not compatible with equilibrium thermodynamics (see Supplementary Note 2 and refs.[11,12] for details). In ref. [11], a p-n-junction was reported that owing to non-ideal behavior shows indications of local stability. Yet recovering of even gentle perturbations takes a long time and large perturbations are found to be detrimental. With the same goal in mind, i.e., trying to avoid dopants, internal p-n transitions have been inspected[13-15] that occur in the bulk of mixed conductors if chemically polarized (e.g., in oxides by applying an oxygen partial pressure gradient). Such hypothetical devices, however, are per se not equilibrium devices, as they involve non-zero electro-chemical potential gradients. A different approach has been taken by utilizing ion conductors. In ref. [16] ionic diodes have been briefly tackled relying on oxygen interstitials (PbO) and vacancies (YSZ, i.e. yttrium doped $ZrO_2$) rather than on p and n. In related studies, $La_2NiO_4$ has been combined with YSZ, or even liquid acid base junctions were investigated[17,18]. Neither are these devices long-time stable nor are they well-working rectifiers.

In this contribution, we - in contrast to previous attempts - refer to genuine *equilibrium space charge effects*, where the "dopant" is mobile and allows for following the bias. The junction under regard is in full electrochemical equilibrium. Hence, we can reversibly switch from electronic depletion (ionic accumulation) to electronic accumulation (ionic depletion). The mobility of the ions in the mixed conductor must be sufficiently large to follow the electric field, but small enough as to not contribute to conductivity effects. As a consequence, the boundary region becomes insulating in the ionic accumulation case while highly conductive in the opposite direction.

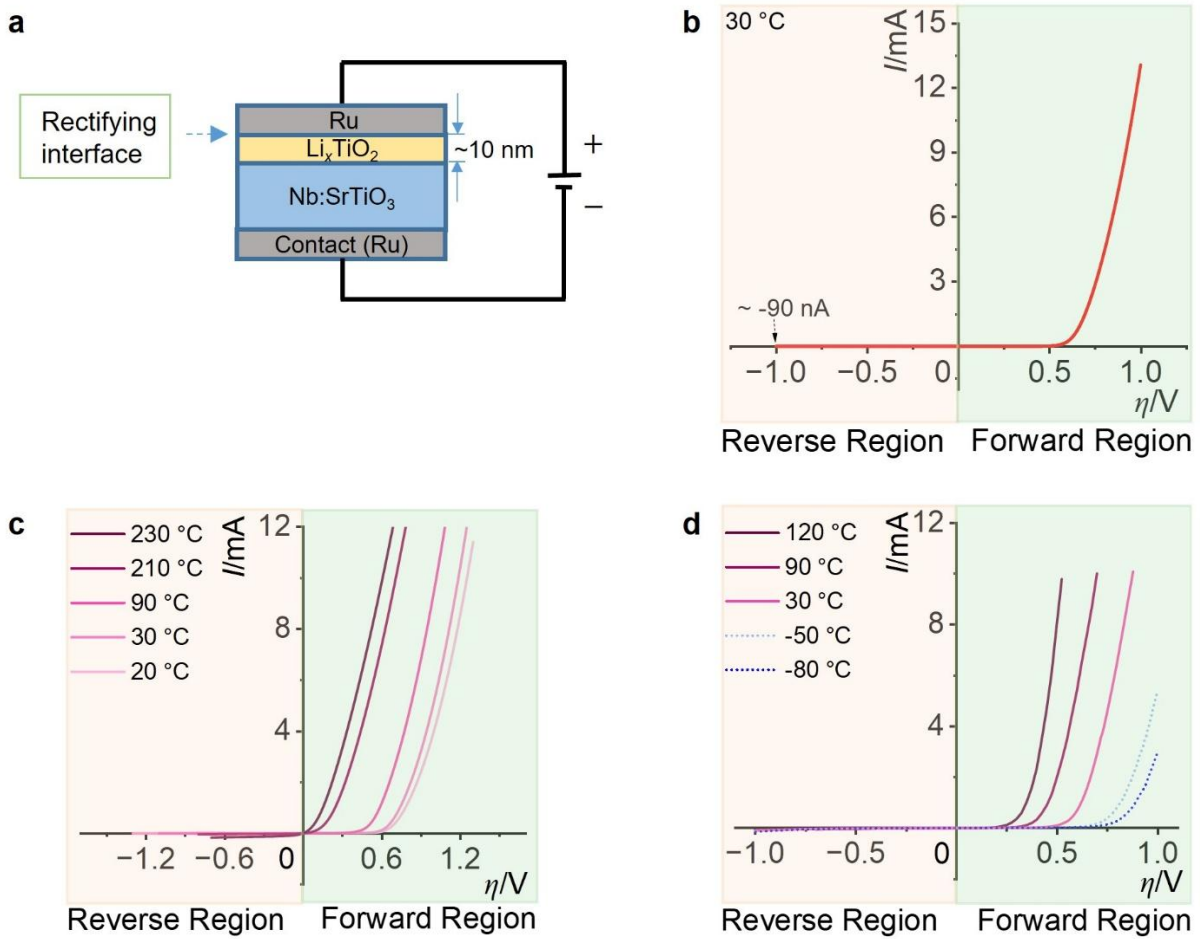

**Fig.1. a** shows the device discussed in the text. **b** displays the rectifying effects for the sample $Li_xTiO_2$ with a Li-content of 0.3. Larger Li-contents depress the reverse current even down to ~1 nA but at the expense of relaxation time. **c** shows the current-voltage curves for the [Li] = 0.35 sample in a temperature window ranging from 20 °C to 230 °C and **d** shows the current-voltage curves for the [Li] = 0.25 sample over a temperature range from -80 °C to 120 °C. While the device works as a Gouy-Chapman diode for room or elevated temperatures, the characteristic changes towards a Schottky-diode at lower temperatures. The device area is 0.8 cm$^2$.

The interface we investigate as a model interface, is Li$_x$TiO$_2$/Ru (Fig. 1a), the highly reversible storage properties of which have been reported in ref. 19 (control experiments on Nb:SrTiO$_3$/Ru and on TiO$_2$/Ru justify our concentrating on the Li$_x$TiO$_2$/Ru interface, see Supplementary Figs. 1 and 2). In the bias-free situation this interface can take up lithium in a job-sharing mode[19-26], whereby Li$^+$ occupies interstitial positions of titania (anatase) whilst the compensating electrons are accommodated on the Ru-side (Li solubility is negligible at operation temperatures). This leads to a significant built-in potential at given Li-contents. As a consequence of the internal electric field, electronic depletion layers occur on the titania side. Bias can increase this depletion or reverse the situation by causing electronic accumulation (Fig. 1b). Owing to the concentration change of also the ions, the time constant of the variation on bias is determined by the chemical diffusion coefficient of Li but also by the size, and can be made very small by using films on the nano-meter scale. In particular at elevated temperatures (Fig. 1c) where the device shows its special advantages over classic rectifiers, response times on the nanosecond scale can be reached (see Extended Data Table 1). The device maintains its functionality also at lower temperatures (Fig. 1d), where the reversible freezing of Li-ion profiles results in a behavior characteristic of a classic Schottky diode.

Such equilibrium space charge layers are termed Gouy-Chapman layers in electrochemistry and widely investigated at interfaces of liquid electrolytes in contact with metallic electrodes. In these electrolytes, the space charge situation is completely made up of ionic charge carriers (anions and cations). A characteristic signature of Gouy-Chapman layers is the occurrence of - in the ideal case symmetric - minima in the capacitance vs. voltage curve, at the point-of-zero charge (flat-band potential). In spite of various efforts, such capacity curves have never been observed for solid electrolytes owing to the large carrier concentrations involved[27,28]. Defined minima have been conversely observed at semiconductor-liquid electrolyte contacts caused by the electron-hole splitting in the semiconductor[29]. All these interfaces are, for obvious reasons, not usable as electronic rectifiers. In this contribution we refer to *mixed conductors*, for which the pair ion/eon dominates the space charge zones and which are expected to exhibit the characteristic features of Gouy-Chapman layers. If the Li content is chosen very high or the temperature very low, the ionic conductivity is much smaller and the Gouy-Chapman behavior turns reversibly into a Mott-Schottky-type of behavior (frozen Li$^+$ profile).

## RESULTS AND DISCUSSION

The requirements for a rectifier to be in full equilibrium in the current-less situation is that not only the electrochemical potentials of the electrons need to be uniform, also the electrochemical potentials of the ionic charge carriers and hence the chemical potentials of the respective neutral components must be without gradients (Fig. 2a, cf. Supplementary Note 2). It is clear that a parent base material with a superimposed dopant profile, as is the case for p-n junctions in electronics, cannot fulfill these criteria, irrespective of the ideality of the situation (Supplementary Note 2). Even for very low diffusion coefficients, such devices must suffer at high temperatures or at extreme miniaturization from drift effects due to local compositional variations driven by the gradients of the electrochemical potentials of the charged dopants (Fig. 2b). Only so-called Gouy-Chapman layers where dopant ions and electrons can follow the electric field, fulfill the thermodynamic requirements.

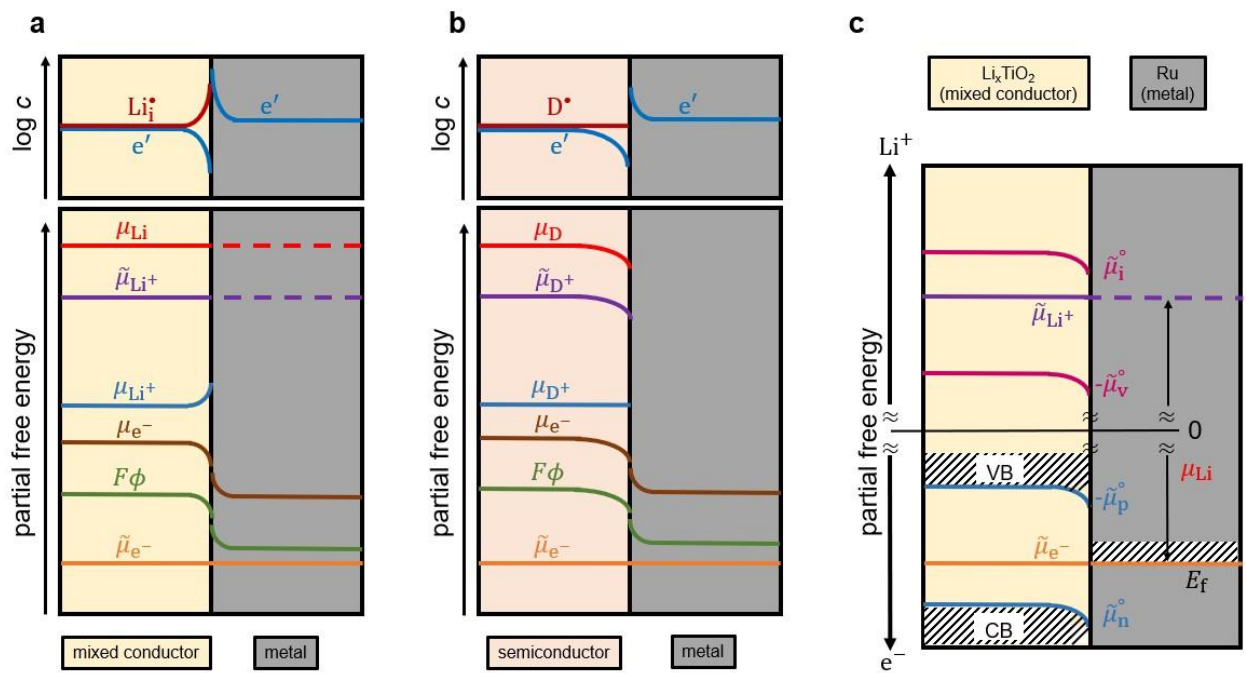

**Fig. 2.** Thermodynamic equilibrium situation of Gouy-Chapman **a** and Mott-Schottky type **b** junctions (sketch of the chemical ($\mu$), electric ($\phi$), and electrochemical potentials ($\tilde{\mu}$) for both cases, see also Supplementary Note 2). *c* in (**a**) and (**b**) stands for concentration. **c**, The same situation as (**a**) but plotted in generalized free energy picture[30] (Note the opposite directions of the partial free energy scales for $e^-$ and $Li^+$). The non-equilibrium situation of conventional semiconductor diodes (**b**) is indicated by the non-zero gradients of the electrochemical potential of $D^+$ (and of the chemical potential of neutral D as component) referring to the current-less situation. The situation under bias is shown in Extended Data Fig. 1.

In order for the device to exhibit a diode function, i.e., being conducting in one and insulating in the opposite overvoltage direction, the counter carrier's conductivity must be negligible, meaning that its mobility must be much smaller than the one of the electronic counter-carrier (but still high enough to quickly follow the applied field). The natural solution is a mixed (ionic/electronic) conductor, where we deal with highly mobile electrons and less but still sufficiently mobile ions. Note that the mobility ($u$) criterion $u(\text{eon}) \gg u(\text{ion}) \gg 0$ does not inflict further restrictions concerning stability, as $u(\text{ion}) \gg 0$ is the necessary criterion for equilibration, and the condition $u(\text{eon}) \gg u(\text{ion})$ only affects the usefulness as an electronic device. The logical candidate is a mixed conductor, in which Li-ions represent the mobile ionic species. In this way, a device can be constructed that is in full thermodynamic equilibrium. Fig. 2a,c shows the necessary and sufficient thermodynamic situation. (See also Supplementary Information.)

The resistance and capacitance of a Gouy-Chapman situation are characterized by distinctive signatures: For ideal conditions (low carrier concentrations), the (perpendicular) resistance is - for large bias - characterized by a 0.5 slope in the semi-logarithmic plot (ln $R^\perp$ versus bias ($\eta$)). (Details are given in the Supplementary Note 3.) In terms of conductivity, the situation changes from a conducting to a rather insulating situation. The capacitance ($C^\perp$ versus $\eta$) exhibits ideally a cosh function with the characteristic minimum at the voltage-of-zero charge ($U_0$) where the profile inverts from an electronic accumulation to an ionic accumulation (and thus an electronic depletion). For large bias, again a slope of 0.5 is expected in the semi-logarithmic plot for ideal conditions (see Supplementary Note 3).

With the help of impedance ($\hat{Z}$) and dielectric modulus ($\hat{M} = j\omega\hat{Z}$) spectra (Fig. 3a,b, see also the Extended Data Figs. 2, 3 and Supplementary Fig. 3), the interfacial contribution could be identified and the interfacial contribution due to Ru/Li$_x$TiO$_2$ separated from other contributions. As Fig. 3c shows, the ideal resistance characteristics are well fulfilled by the Ru/nano-Li$_x$TiO$_2$ junction for not too low bias ($\eta$) values for which saturation effects become important. The capacitance curve (Fig. 3d) indeed shows a minimum at -0.5 V which can be affiliated with the point-of-zero charge. Interestingly the entire curve is approximately ideal for voltages for which the metal side is expected to be positively charged while less ideal on the side where the metal is negatively charged. This is understandable, because Ru as a metal – though a noble metal – is not an ideal electron acceptor.

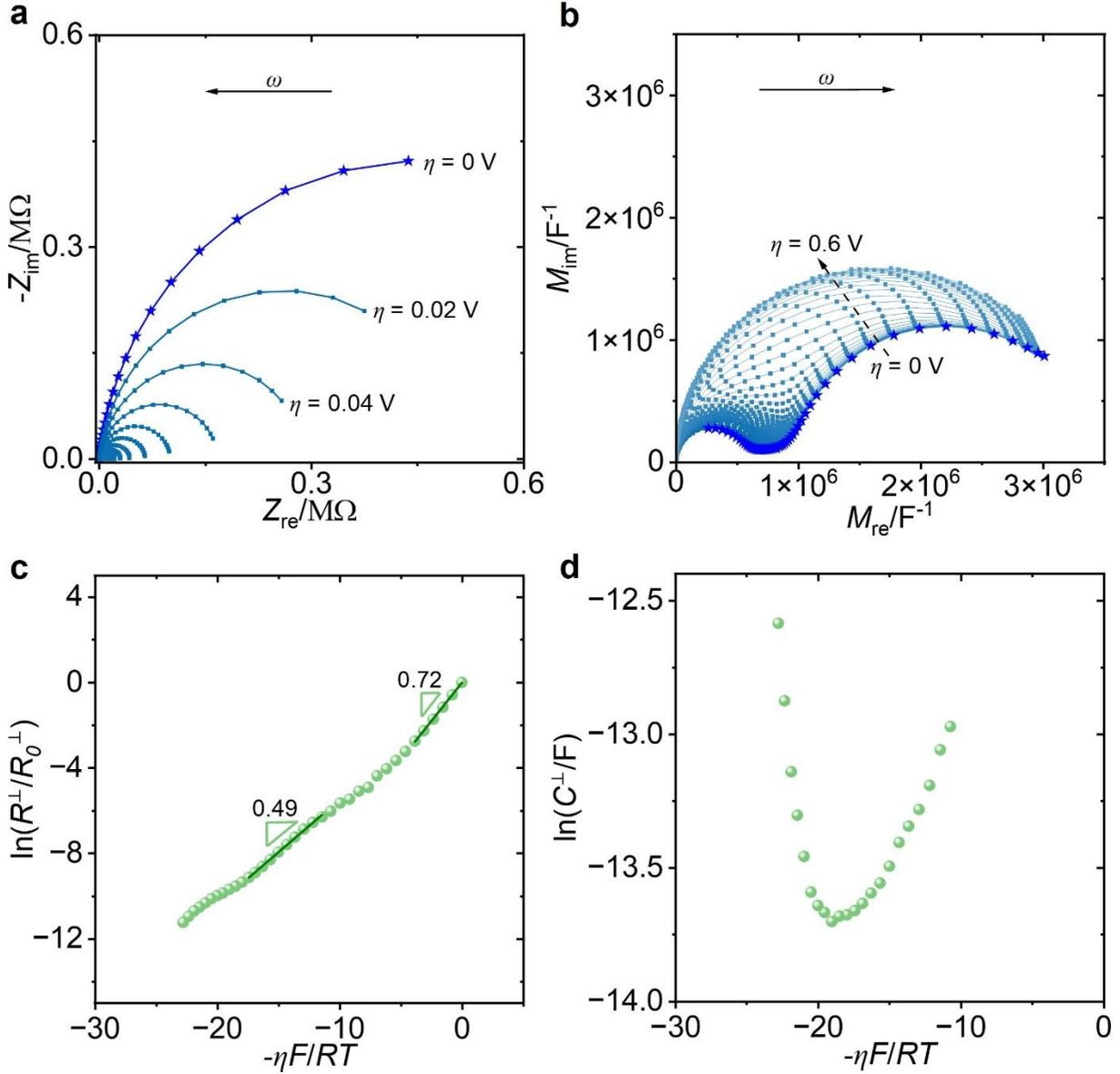

**Fig. 3. Impedance measurement perpendicular to the interface. a**, **b**, Impedance and dielectric modulus plots at 30 °C for the cell shown in Fig. 1a. In the modulus plots the two contributions (bulk and interface) become clearer. **c**, Interfacial resistance as a function of bias ($\eta$). The middle region shows the ideal slope of 0.5, while at lower bias ($\eta$) saturation effects come into play. **d**, The capacitance curve exhibits a clear minimum displaying the point of zero charge (pzc) of – 0.5 V for this special Li content (~0.2). The behavior left from pzc is almost ideal, while the right-hand side part appears to be influenced by electronic saturation on the Ru-side. (Extended Data Fig. 2 shows analogous curves for various Li-concentrations).

Even though addressed below again, we will mention further advantages of our device: More suitable threshold voltages, possibility of tuning the characteristics by varying the Li content (*x*) (see Fig. 4a-c) and the fact that no high temperatures are needed for the preparation process. It is of special worth that the onset voltages can be varied over a wide range (from a low voltage range of 0.2 - 0.3 V typical for Schottky diodes to a 0.5 - 0.7 V typical for p-n junction rectifiers) (Fig. 4c). Even a very low voltage range (≳ 0.1 V) is addressable which makes the device suitable for low power application (Extended Data Fig. 4). This range could not be made accessible for Schottky diodes. (In-situ tuning of the onset voltage was also reported for conventional Schottky diodes if the metal is replaced by conducting polymers[31].). Conventionally, forward voltage variations required utilizing diverse material systems with tailored barrier heights (see Supplementary Table 1).

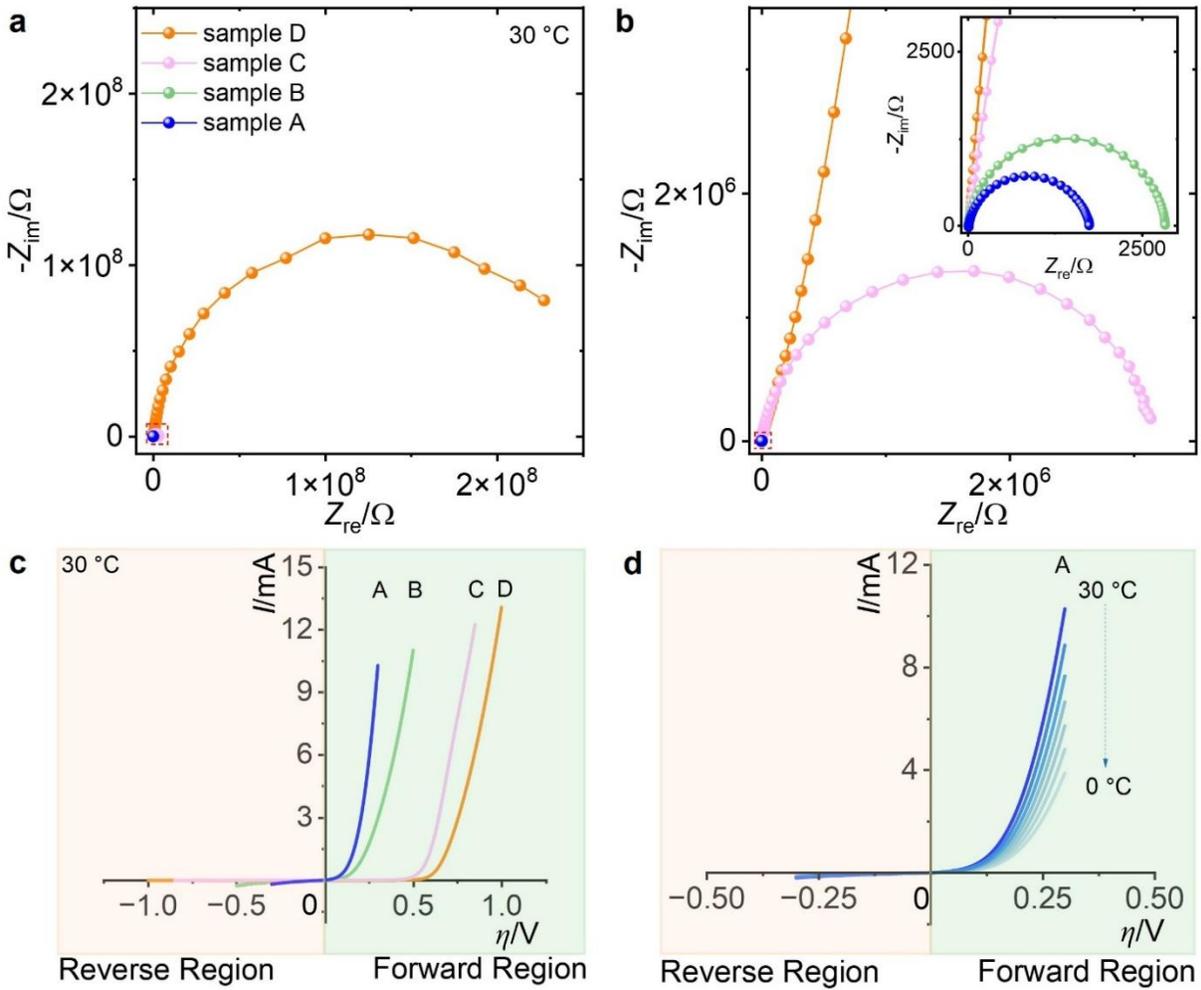

**Fig. 4. Characteristics tuning. a**, **b**, Impedance plots at different resolutions ((**b**) magnifies the region around zero point in (**a**) – red dashed box and the inset in (**b**) magnifies the region around zero point in (**b**)) and **c** current-voltage curves for 4 different Li-contents (A: 0.02, B: 0.03, C: 0.2, and D: 0.3). The impedance plots display the significant interfacial resistance variations if the Li content is varied. Whilst the temperature for (**a - c**) was 30 °C, **d** shows the effect of lowering the temperature on the current-voltage curve for the [Li] = 0.02 (sample A).

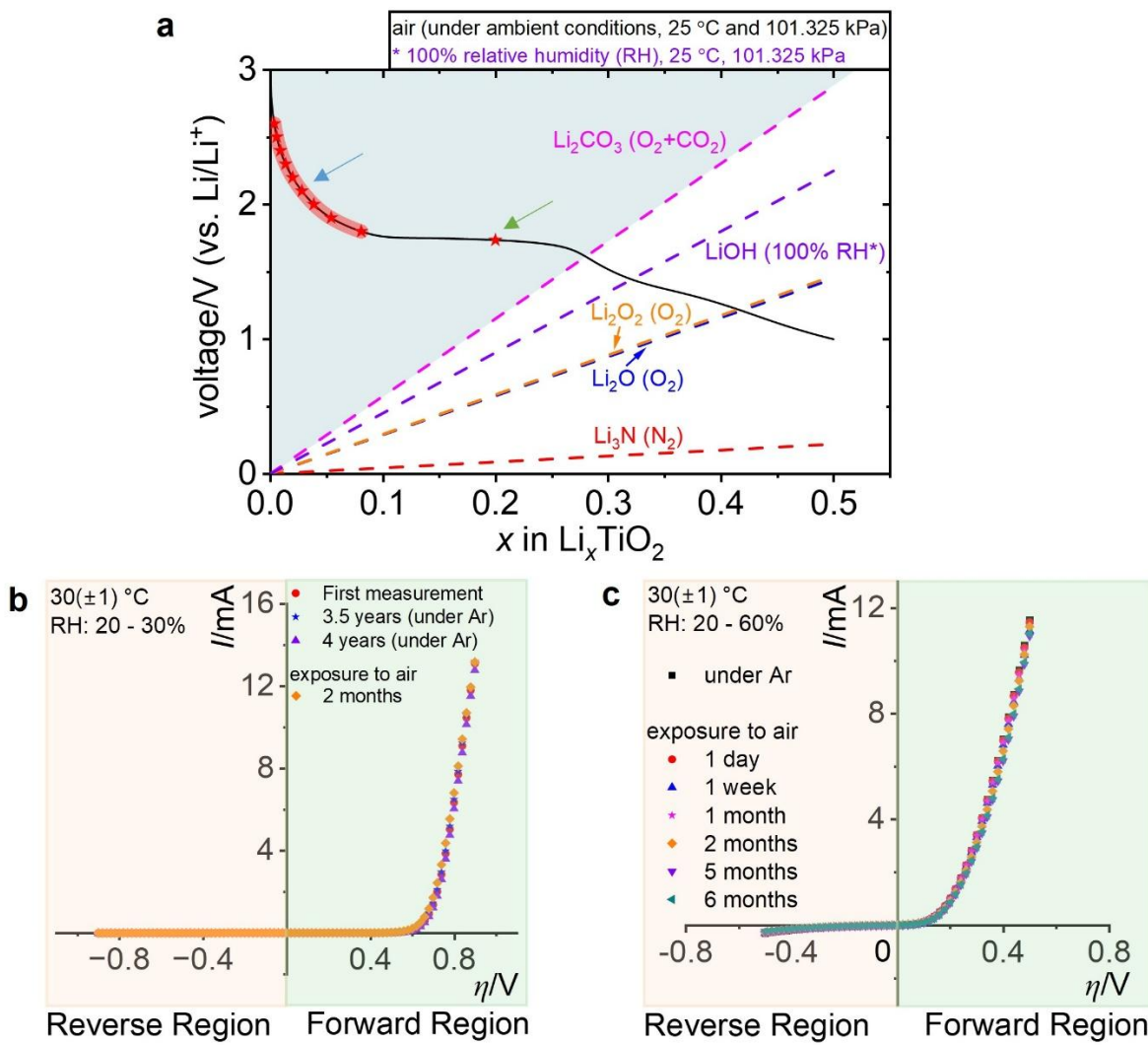

**Fig. 5. Thermochemical stability of Li$_x$TiO$_2$. a**, The thermochemical stability under ambient conditions is guaranteed for Li-concentrations the OCV of which exceeds the reaction equilibrium lines (light blue region), as outlined in the Supplementary Note 4. We show that our master example is not only inherently stable, but also environmentally stable. Collecting the Gibbs free enthalpy values of the virtually involved compounds and compare them with the OCV data vs. Li/Li$^+$ we can derive that for air conditions at 25 °C, $x < 0.25$ is sufficient to guarantee that no reaction to Li$_2$O (or Li$_2$O$_2$) or LiOH or Li$_2$CO$_3$ or Li$_3$N is possible. For higher $x$ values one may seal the device in the same way as it is common for Li batteries. **b** shows that even waiting times as long as 50 months do not change the current-voltage relation. The sample ($x$~0.2, indicated by the green arrow in (**a**)) has been kept under Ar for 4 years and exposed to air for 2 months. **c** further shows that this precaution is not necessary (in agreement with (**a**)). The current-voltage relation under Ar and (dry or humid (relative humidity (RH): 20 – 60 %)) air do not differ measurably (for the Li content indicated by the blue arrow in (**a**)). The devices indicated in the red shaded region (**a**) show good functionality.

The impressive long-time stability is particularly obvious from Fig. 5 where the device with $x = 0.2$ was kept for 4 years at 25 °C and did not show any perceptible changes when compared to the initial curves, i.e., showing no signs of drift effects. The sample had been kept in the glove box (under Ar, $O_2 < 0.1$ ppm and $H_2O < 0.1$ ppm), which in view of the thermodynamic stability under ambient conditions (see Fig. 5a and Supplementary Note 4) turned out to be an unnecessary precaution ($x < 0.25$). The current-voltage relation showed no measurable drift after the sample had been exposed to air for 2 months (Fig. 5b, sample 1) and 6 months (Fig. 5c and Supplementary Fig. 4, sample 2).

The temperature stability is indicated by the impedance studies in Extended Data Fig. 5, where the impedance spectra recorded during cooling and subsequent heating at each temperature are identical (−196 °C to 30 °C). The practical operation temperature is restricted to $T < 250$ °C where the Li solubility in Ru is no longer negligible, thus changing built-in potential, current on-off ratio and response time (see Fig. 1c).

The operational functioning as rectifier is best shown by Fig. 6 where voltage steps in different directions have been applied and the current response was followed.

The fact that we change both the $Li^+$ and the $e^-$ profiles means that the bias changes locally the Li content. Hence, the chemical diffusion of Li in the space charge zone determines the response time of our device. On the nanoscale, this relaxation time can be less than 30 ms at room temperature or above as depicted by Fig. 6. The ms response can be accelerated to a ns response at elevated temperatures, where the advantage of our device would be more pronounced (Extended Data Table 1). Note that this time behavior characterizes the relaxation towards equilibrium and should not be confused with an irreversible drift effect that would be characteristic for the long-time behavior of a non-equilibrium device. Our device quickly returns to the initial value when the bias is turned off (Supplementary Fig. 5). Interestingly, we can switch from the Gouy-Chapman equilibrium device to a quasi-Schottky device by drastically reducing temperature (see Fig. 1d) (We do not refer to a classic Schottky device where the dopant profile is horizontal, rather we address a frozen Gouy-Chapman profile. The electrochemical characteristics of such a situation, which in fact should be typical for most frozen cases, will be discussed elsewhere[32].) In this frozen state the relaxation time of the ions is negligibly small. If departing from such "freezing" condition, one returns to a Gouy-Chapman situation.

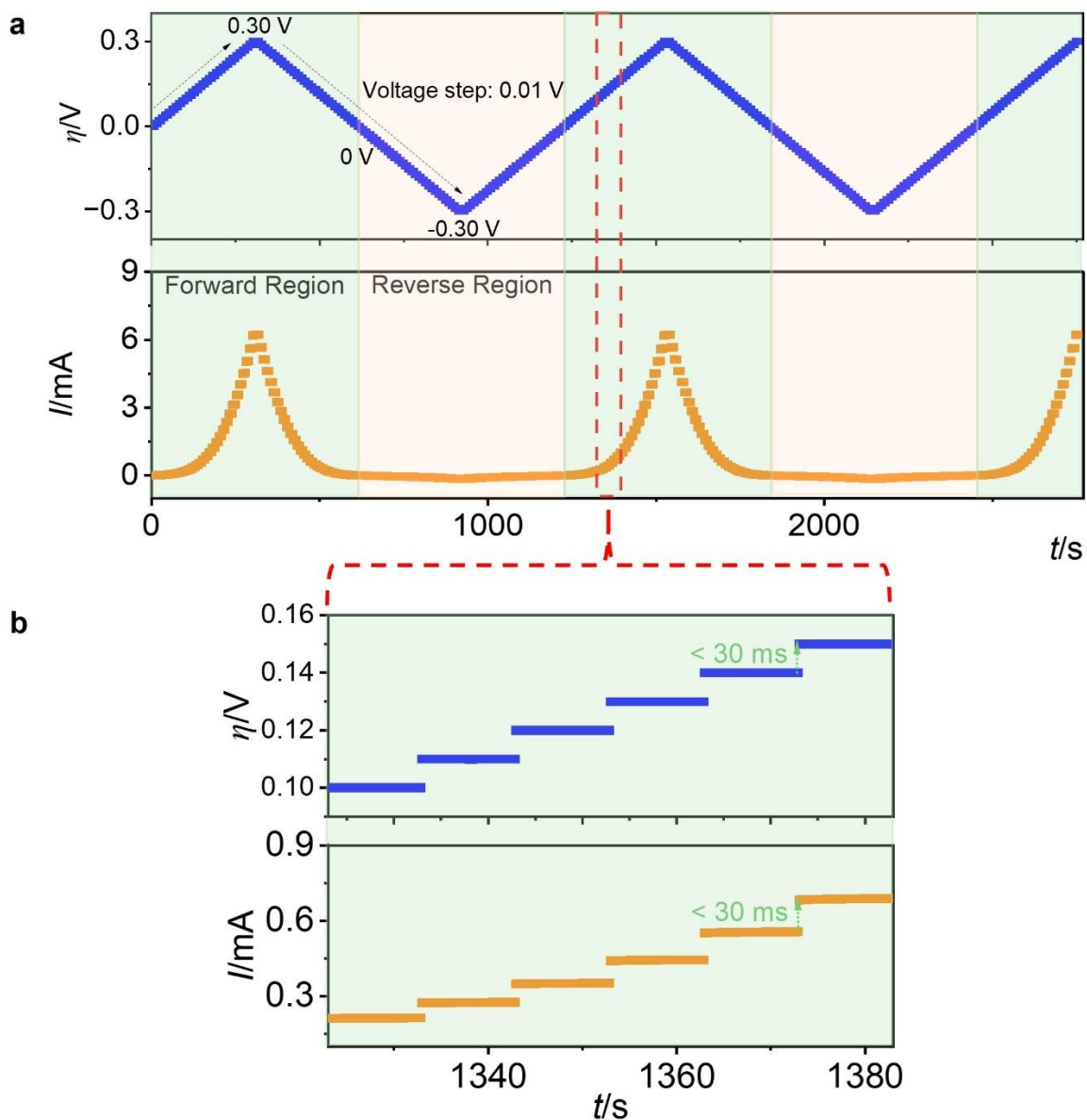

**Fig. 6. a**, Current response under voltage step. Voltage steps (0.01 V/step) in forward and reverse directions are applied. The sample ($Li_xTiO_2$, $x\sim0.025$) thickness is 12 nm. **b** magnifies the region indicated in (**a**) (red dashed box). The relaxation time is shorter than 30 ms.

Notwithstanding the fact that the ions' long-range motion is suppressed at high frequencies, our device should also work at such frequencies (Supplementary Fig. 6). Here, the rectifying mode will again be similar to that of a Schottky diode with however the advantage that at rest it will relax rather quickly, while the Schottky diode will exhibit the above-mentioned drift. Again, one comes back to the Gouy-Chapman situation by reducing the frequency.

As already mentioned, it is a further substantial advantage of our device that the preparation (unlike conventional diodes) does not need a high-temperature step; it is also much less expensive and much more gentle than e.g., ion implantation[33-38]. What is more, the "doping" (i.e., Li-)content can be precisely and elegantly varied electrochemically. Beyond that, the Li-solubility is so high (as compared to the band gap) that intrinsic effects do not become significant even at high temperature (this is in contrast to Si-based devices)[1-5]. In this way, the onset voltage (see Fig. 4) can be varied and adjusted to the wanted situation. Beyond that, the thin-film preparation procedure allows for easy and precise control of the dimensions. Under Gouy-Chapman conditions, the screening length is on the order of the lattice parameter, enabling nanometer-scale device thickness thus enabling providing a straightforward pathway for miniaturization. This is in contrast to classic devices which rely on the much wider depletion regions of a Mott-Schottky situation (e.g., the classic p-n junction and Schottky diode) and generally work on the micrometer scale.

The general concept of nanoionic rectifiers presented here, can be extended to a wide range of other material systems (e.g., Ru/Li$_x$Nb$_2$O$_5$, see Extended Data Fig. 6).

CONCLUSION

In summary, we report on a nanoionic rectifier using the behavior of equilibrium space charge zones in mixed conductors which shows absolutely no drift even under conditions where previous devices must fail due to the finite dopant mobility. The Ru/Li$_x$TiO$_2$ interface fulfills these requirements very satisfactorily. The relaxation time on voltage changes is given by the chemical diffusion coefficient of Li and thus determined by size, Li-content and thickness. It can, as shown, be easily reduced to negligibly small values. Another advantage is that the series resistance is small, and the onset voltage can be adjusted by varying the Li-content and thus the device can also be used at low power. Generally, the Li-content can be much higher than typical doping contents in e.g., Si, so that we refer to extrinsic conditions even at high temperatures. Straightforward preparation without the necessity of high-temperature treatment, the in-situ variation of size as well as the possibility of electrochemically varying the "doping" content at room temperature, are further important practical assets. The device can be used even at very high frequencies (or at very low temperature) where ionic motion is frozen; it will then work as a usual space charge diode. The major task of our investigation was to show how to design *equilibrium* rectifiers using mixed conductors. In this way, a path is opened to novel rectifiers that are based on Gouy-Chapman layers rather than Mott-Schottky layers. Sodium (Na)-based mixed conductors would be a realistic alternative with the perspective of even higher operation temperatures. The use of Ag or Cu based mixed conductors might lead to an even quicker response owing to higher diffusivities but potentially at the expense of chemical stability. Certainly, also the precious Ru metal may be replaced by a different stable electronic conductor that shows negligible Li solubilities.

**ACKNOWLEDGEMENTS**

We thank R. Merkle, Y. Zhang, D. Moia, and R. Usiskin for helpful discussions; F. Kaiser, C. Gan, H. Hoier, A. Sorg, U. Klock, A. Fuchs, U. Traub, K. Küster, T. Reindl, and M. Hagel for technical support; and H. Klauk for critically reading the manuscript.

**Funding:**

This work received no external funding.

**Author contributions:**

C.X. and J.M. conceived the project and designed the experiments. J.M. supervised the project. C.X. executed the materials preparation, materials characterization, electrochemical, and transport experiments. C.X. and J.M. analyzed the data and were responsible for the theoretical treatment, manuscript preparation, and manuscript revision.

**Extended Data Figures and Table**

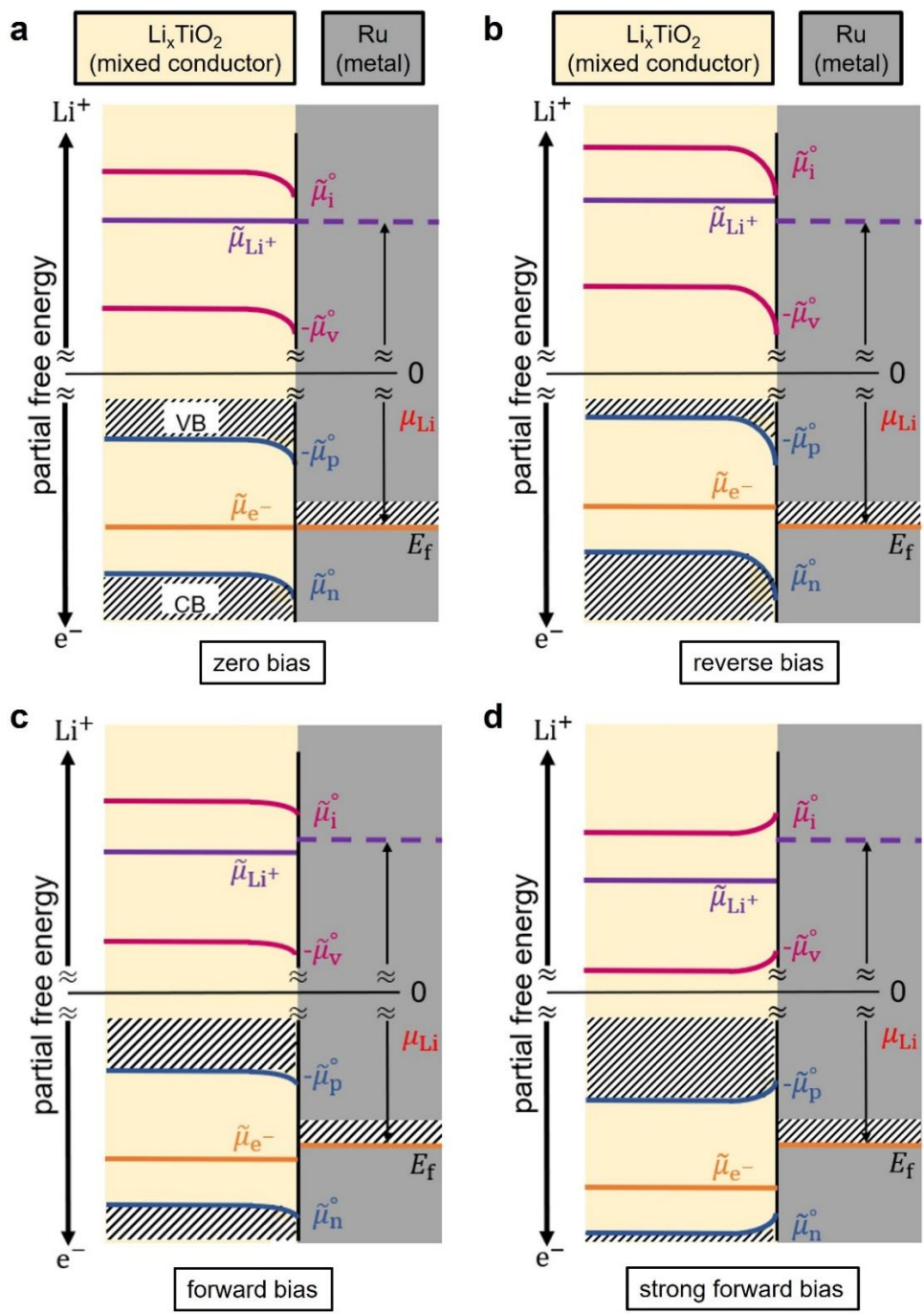

**Extended Data Fig. 1. Generalized (free) energy level diagram of mixed conductor (e.g., Li$_x$TiO$_2$) and a metal (e.g., Ru) contact under different biasing conditions**: **a**, Equilibrium state. **b**, Reverse bias. **c**, Forward bias. **d**, Strong forward bias. Note that the electronic level diagram is plotted upside down. For simplicity the chemical potential of Li is assumed to be uniform (even in the metal; in praxi this is not important as the Li-solubility in the metal is negligible). As the Li conductivity in the metal is assumed to be minute, Li-equilibrium in the metal should anyway not be relevant for our discussion. For simplicity we also neglect stochiometric polarization.

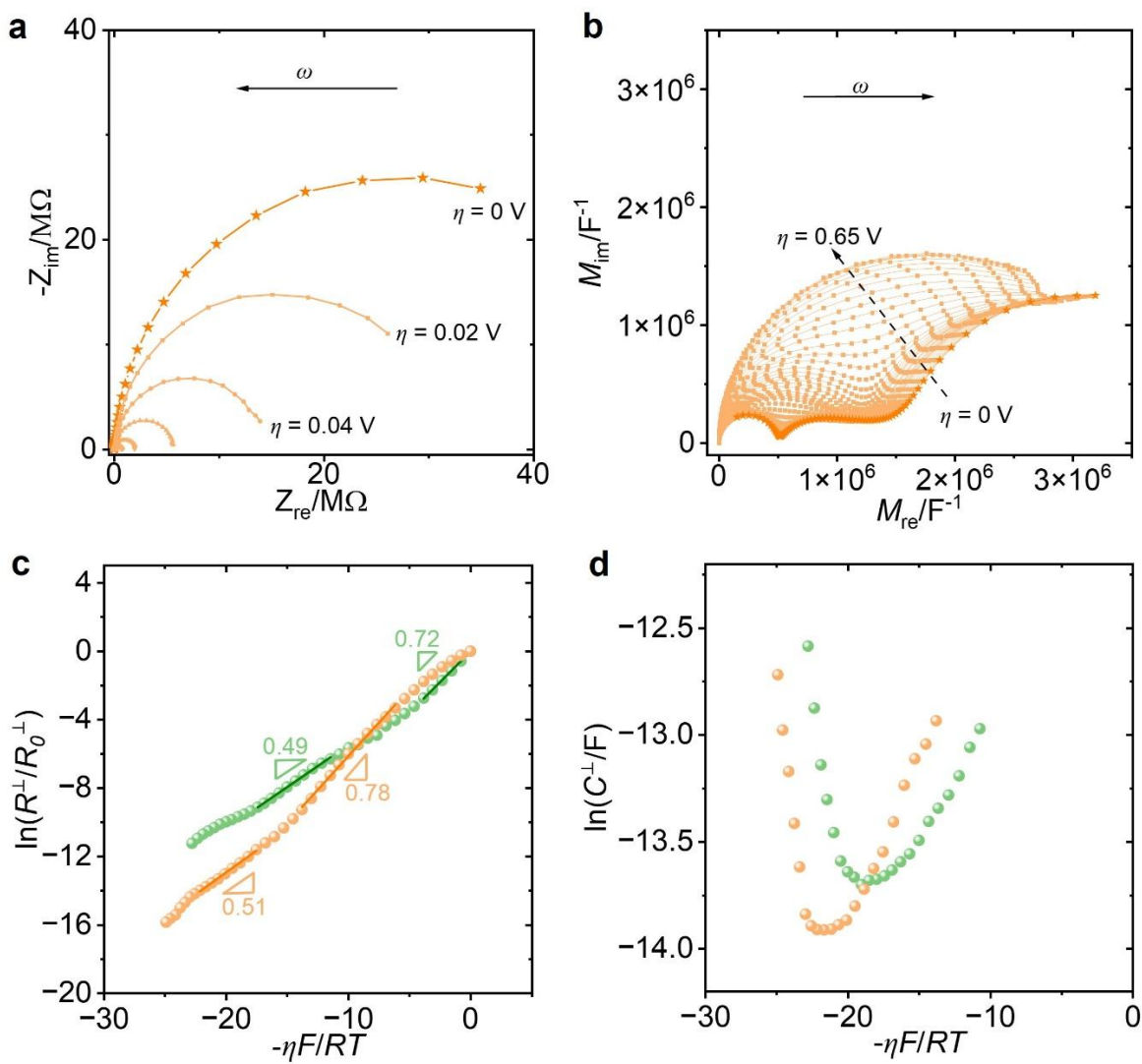

**Extended Data Fig. 2. Impedance measurement perpendicular to the interface**. **a,b**, Impedance and dielectric modulus plots at 30 °C for the cell shown in Fig. 1a. In the modulus plots the two contributions (bulk and interface) become clearer. **c**, Interfacial resistance as a function of bias ($\eta$) for the samples with different Li concentrations ([Li]~0.25 (orange) and [Li]~0.20 (green, see also Fig. 3 in the main text)). The middle region shows the ideal slope of 0.5, while at lower bias ($\eta$) saturation effects come into play. **d**, The capacitance curve exhibits a clear minimum displaying the point of zero charge (pzc) to be – 0.57 V for Li content (~0.25) and – 0.5 V for Li content (~0.20). The behavior left from pzc is almost ideal, while the right-hand side part appears to be influenced by saturation on the Ru-side.

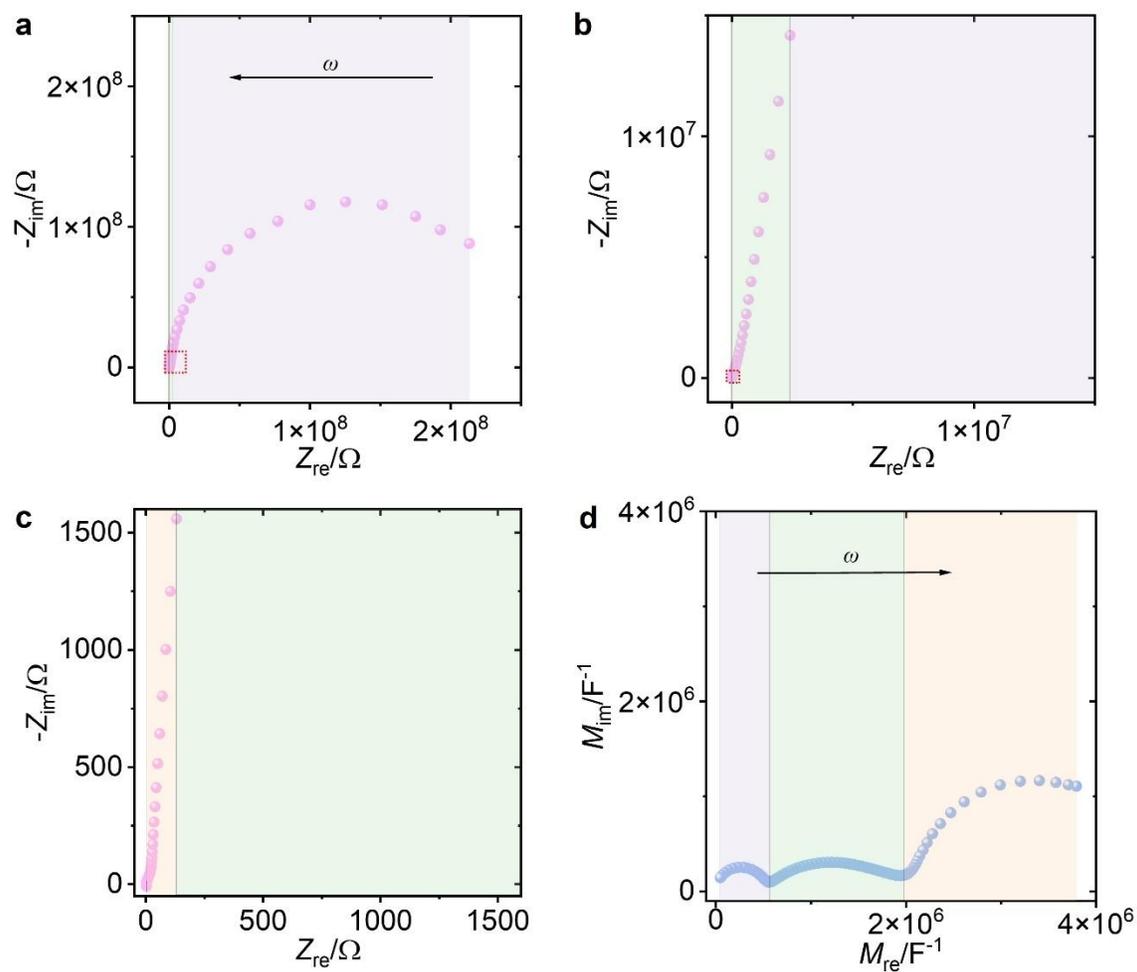

**Extended Data Fig. 3.** Impedance (**a-c**) and dielectric modulus (**d**) plots for the device shown in Fig. 1a, displayed at different magnifications ((**b**) magnifies the region around zero point in (**a**) – red dashed box and (**c**) magnifies the region around zero point in (**b**)). The bulk contribution and interfacial contribution could be identified.

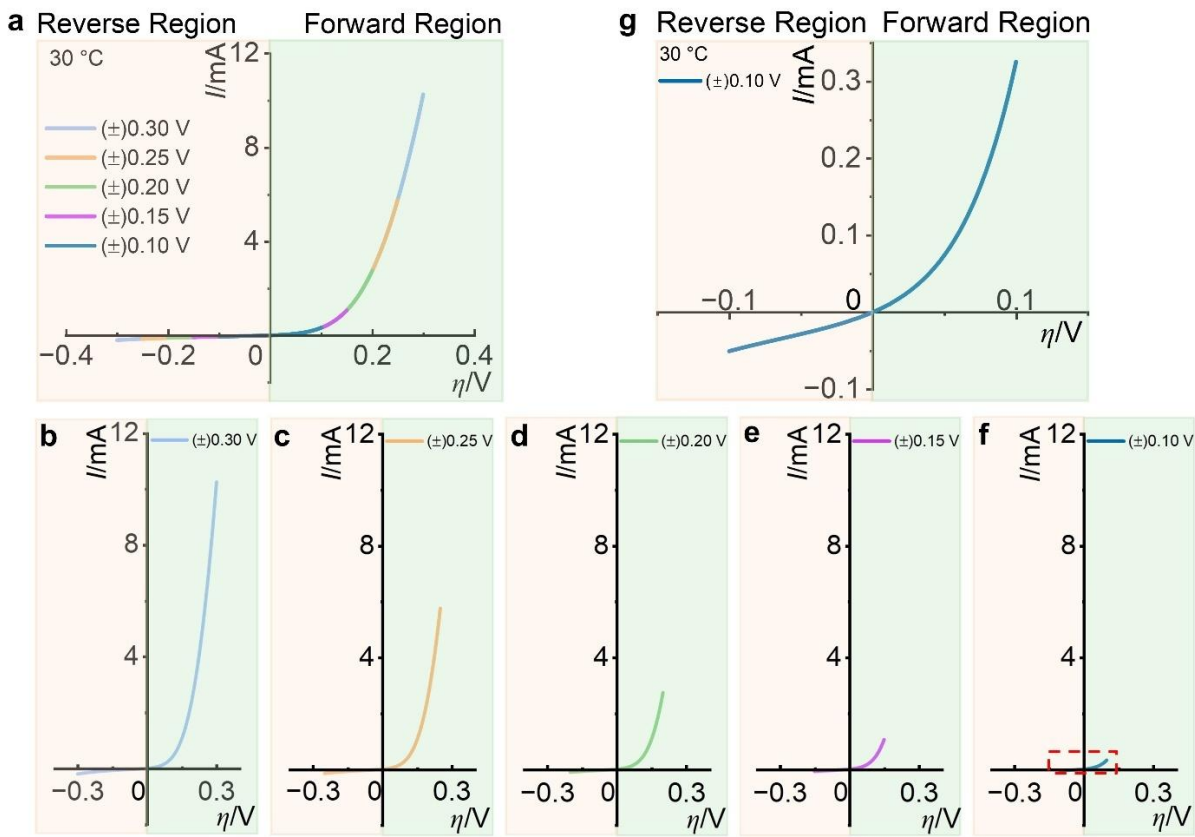

**Extended Data Fig. 4.** Current–voltage curves of the sample with a Li content of 0.03 measured at 30 °C under different bias ranges **a**. The device exhibits good stability and reversibility across all applied voltage ranges. **b-f** present the corresponding current-voltage curves from (**a**) separately: (**b**) (±)0.30 V; (**c**) (±)0.25 V; (**d**) (±)0.20 V; (**e**) (±)0.15 V; and (**f**) (±)0.10 V. **g**, Enlarged current-voltage curve corresponding to (**f**) (the range indicated by red dashed box in (**f**)). These results highlight one of the key advantages of the device - its suitability for low-power applications (A voltage of ~0.10 V is sufficient to maintain the rectification function.).

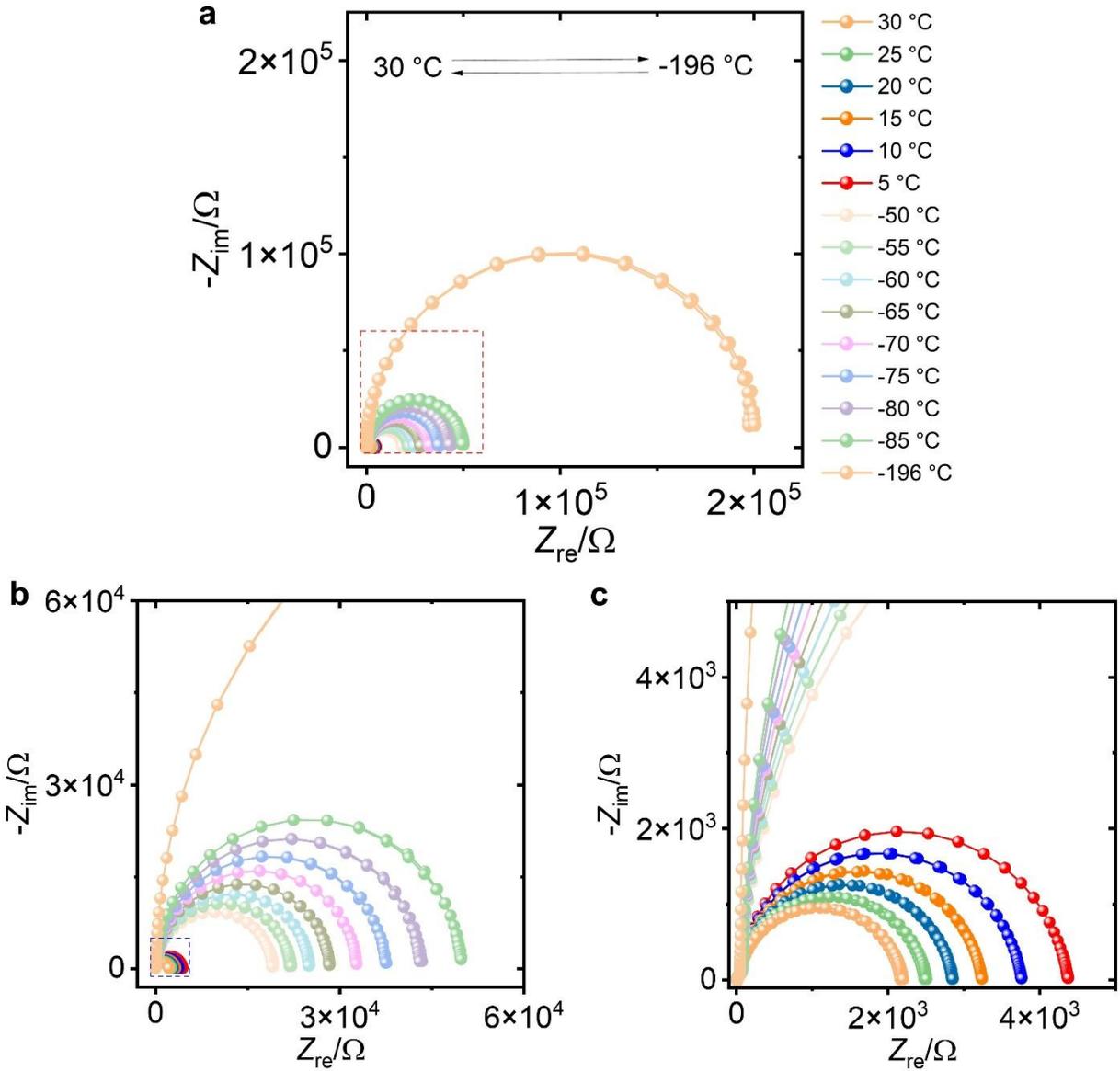

**Extended Data Fig. 5. Temperature stability measurements**. **a-c**, Impedance spectra of the sample with a Li content of 0.03 measured at temperatures ranging from −196 °C to 30 °C, shown at different magnifications in (**a-c**). (**b**) magnifies the region highlighted by the red dashed box in (**a**), while (**c**) further enlarges the region near the origin indicated by the blue dashed box in (**b**). The impedance spectra recorded during cooling and subsequent heating at each temperature overlap, indicating excellent thermal stability of the sample.

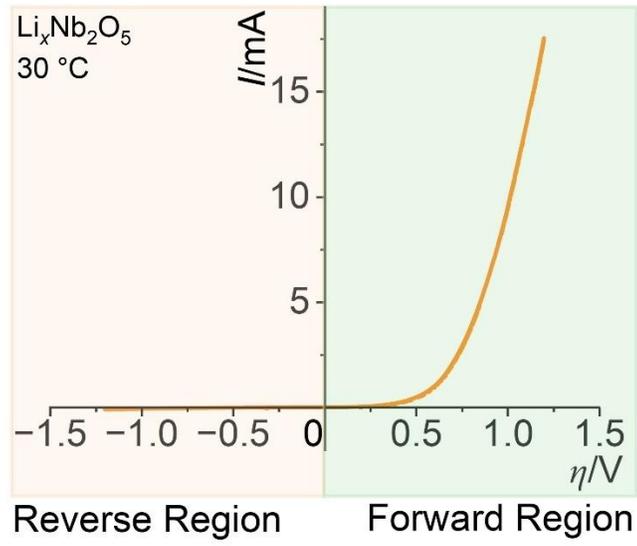

**Extended Data Fig. 6.** Current–voltage curve of lithiated $Nb_2O_5$ thin films measured at 30 °C.

**Extended Data Table 1.** Estimated equilibration times for Li diffusion in anatase at different temperatures (30 °C, 400 °C, and 700 °C). The diffusion lengths $L$ (corresponding to the sample thicknesses) are taken as 6, 12, and 24 nm, respectively. Taken from our experiment, a room-temperature chemical diffusion coefficient of Li in anatase can be estimated as $D^\delta = 5 \times 10^{-11}$ cm² s⁻¹ (sample size: 12 nm; equilibration time: 30 ms). The activation energy is assumed to be 0.5 eV, a typical value for electronic conductor with comparable amounts of ionic and electronic charge carriers[42-48]. The calculated $\tau$-values for space charge control[14] are lower assuming a Debye-length smaller than the lattice parameter.

| L/nm | $\tau^\delta_{eq}$ (30 °C) $D^\delta = 5 \times 10^{-11}$ cm² s⁻¹ | $\tau^\delta_{eq}$ (400 °C) $D^\delta = 2 \times 10^{-6}$ cm² s⁻¹ | $\tau^\delta_{eq}$ (700 °C) $D^\delta = 2.5 \times 10^{-5}$ cm² s⁻¹ |
|---|---|---|---|
| 6 | 7 ms | 180 ns | 14.5 ns |
| 12 | 29 ms | 720 ns | 58 ns |
| 24 | 115 ms | 2880 ns | 230 ns |